# Virtual Astronomy, Information Technology, and the New Scientific Methodology

S. G. Djorgovski, *California Institute of Technology*

*Abstract*—All sciences, including astronomy, are now entering the era of information abundance. The exponentially increasing volume and complexity of modern data sets promises to transform the scientific practice, but also poses a number of common technological challenges. The Virtual Observatory concept is the astronomical community's response to these challenges: it aims to harness the progress in information technology in the service of astronomy, and at the same time provide a valuable testbed for information technology and applied computer science. Challenges broadly fall into two categories: data handling (or "data farming"), including issues such as archives, intelligent storage, databases, interoperability, fast networks, etc., and data mining, data understanding, and knowledge discovery, which include issues such as automated clustering and classification, multivariate correlation searches, pattern recognition, visualization in highly hyperdimensional parameter spaces, etc., as well as various applications of machine learning in these contexts. Such techniques are forming a methodological foundation for science with massive and complex data sets in general, and are likely to have a much broather impact on the modern society, commerce, information economy, security, etc. There is a powerful emerging synergy between the computationally enabled science and the science-driven computing, which will drive the progress in science, scholarship, and many other venues in the 21st century.

*Index Terms*—Astronomy; Data management; Information technology; Knowledge acquisition; Knowledge representation; Scientific visualization.

## I. INTRODUCTION

WE are on the cusp of the second stage of the information technology revolution. The past 2 or 3 decades have been dominated – and the world transformed – by the advent of increasingly more powerful, less expensive, and ubiquitous computing, and the appearance of the World Wide Web and related technologies. But the rise of information technology (IT) has also generated a whole new set of challenges: the world is drowning in a tidal wave of data, which increase exponentially both in the volume and complexity. Coping with the challenge of effective utilization of information-rich data sets forms the battleground of the second stage of this new scientific and industrial revolution.

One could consider the modern computation and communication hardware mainly as the enabling technology for the information processing (in the broadest sense) software technology, which in itself is essentially an enabling technology for the new methodology of science, engineering, and operating practices of modern industry, commerce, and indeed all fields of contemporary human endeavor.

In order to illustrate these concepts, in this paper we consider a specific example of the development of information-rich astronomy, and the technological responses our community has devised in order to cope with the challenge and opportunity posed by massive and complex data sets. Similar issues apply to many other sciences, with potential utility and applications well beyond the academia.

## II. THE VIRTUAL OBSERVATORY CONCEPT

The data volume in astronomy grows exponentially, with a doubling time scale of ~ 1.5 years [1]. The similarity with the Moore's law is no accident: the same technology (mainly VLSI) is also responsible for the growth of astronomical detectors (e.g., CCDs and other imaging arrays) and computer-based data systems used to gather the observations. There are now (~ early 2005) estimated 0.5 – 1 Petabytes (PB) in various accessible astronomical archives and data depositories [2], with at least a comparable amount of legacy data yet to be ingested; this does not include planetary astronomy and space physics. The current data generation rate in astronomy is about 1 Terabyte (TB) per day (mainly night, actually). Note that both data volume and data rate grow exponentially.

It is not only the data volume which is increasing, but also data complexity, homogeneity, and overall quality. Most of the data come from wide-field surveys, which now span a full range of wavelengths, from radio to γ-rays, typically generating tens or hundreds of TB, detecting millions to billions of sources (stars, galaxies, etc.), and measuring tens to hundreds attributes per source. Data come usually in some multidimensional form as images, spectra, data cubes, time series, etc. We are building a more complete, fully panchromatic picture of the physical universe. Modern astrophysical theory is also a prolific producer of comparable

An invited review, to appear in IEEE Proc. of CAMP05, *Computer Architectures for Machine Perception,* eds. V. Di Gesu & D. Tegolo, in press (2005)

Manuscript submitted on March 7, 2005. This work was supported in part by the U.S. National Science Foundation grants AST-0122449, AST-0326524, AST-0407448, DMS-0101360, NASA contract NAG5-9482, and the Ajax Foundation.
S. G. Djorgovski is with Division of Physics, Mathematics, and Astronomy, and with Center for Advanced Computing Research, California Institute of Technology, MS 105-24, Pasadena, CA 91125, USA; telephone +1 626 395 4415; e-mail: george@astro.caltech.edu.

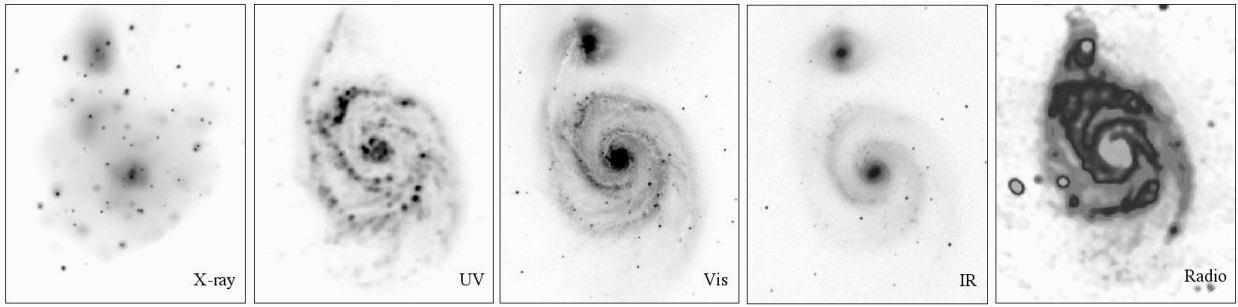

**Figure 1.** A multi-wavelength view of the nearby spiral galaxy M51, illustrating the different appearance of complex astrophysical objects in different wavelength regimes. From the left: an X-ray image from the *Chandra* observatory; a UV image from the *GALEX* satellite; a visible light image; a near-IR image from the *2MASS* survey; and a 6 cm radio continuum image from the Effelsberg telescope.

volumes of highly complex data generated by numerical simulations. Much of the new science comes from federation of such massive and already complex data sets, thus leading to even greater data handling and analysis challenges. The only hope of understanding complex astrophysical phenomena, e.g., star and planet formation, galaxy and large-scale structure formation and evolution, explosions of supernovae, etc., is in having sufficiently complex and comprehensive data and theoretical simulations, and being able to combine them is a scientifically valid and effective manner.

Yet, it is clear that our understanding of the universe does not double every year and a half. We are not yet utilizing the full information content of these rich (and usually expensive) data sets. There seems to be a methodological bottleneck in the conversion of masses of data bits into actual knowledge.

The Virtual Observatory (VO) concept is the astronomical community's response to these challenges and opportunities. VO is an emerging, open, web-based, distributed research environment for astronomy with massive and complex data sets. It assembles data archives and services, as well as data exploration and analysis tools. It is technology-enabled, but science-driven, providing excellent opportunities for collaboration between astronomers and computer science (CS) and IT professionals and statisticians. It is also an example of a new type of a scientific organization, which is inherently distributed, inherently multidisciplinary, with an unusually broad spectrum of contributors and users.

The concept was defined in late 1990's through many discussions and workshops, culminating in a significant endorsement by the U.S. National Academy's influential "decadal survey" report [3], and a white paper [4] and other contributions to the first major conference on the subject [5]. It was then further refined by the U.S. National Virtual Observatory Science Definition Team, whose report provided the most comprehensive description of the concept and the background up to that point [6]. More international conferences followed [5,7,8], and a good picture of this emerging field can be found in papers contained in their proceedings. VO projects have been initiated world-wide, with a good and growing international collaboration between various efforts [9,10]. More links can be found on the author's website [11].

While any individual function envisioned for the VO can be accomplished using existing tools, e.g., federating a couple of massive data sets, exploring them in a search for particular type of objects, or outliers, or correlations, in most cases such studies would be too time-consuming and impractical; and many scientists would have to solve the same issues repeatedly. VO would thus serve as an enabler of science with massive and complex data sets, and as an efficiency amplifier. The goal is to enable some qualitatively new and different science, and not just the same as before, but with a larger quantity of data. We will need to learn to ask different kinds of questions, which we could not hope to answer with the much smaller and information-poor data sets in the past.

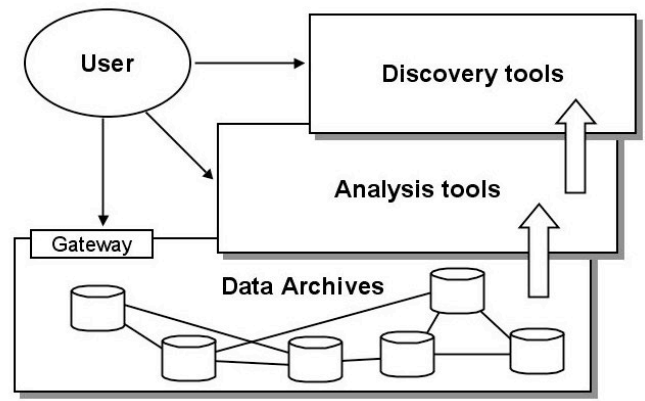

**Figure 2.** A conceptual architecture of a VO, from a science user's viewpoint. A user should be able to discover the available data for their study, which generally reside in distributed archives, federate them, and pipe the output into a set of DM/KDD data analysis and discovery tools, which may be implemented as web services, and may involve use of AI and machine learning tools, coupled with sophisticated visualization environments.

Looking back at the history of astronomy we can see that technological revolutions lead to bursts of scientific growth and discovery. For example, in the 1960's, we saw the rise of radio astronomy, powered by the developments in electronics (which were much accelerated by the radar technology of the World War II and the cold war). This has led to the discovery of quasars and other powerful active galactic nuclei, pulsars, the cosmic microwave background (which firmly established the Big Bang cosmology), etc. At the same time, the access to space opened the fields of X-ray and γ-ray astronomy, with an equally impressive range of fundamental new discoveries: the very existence of the cosmic X-ray sources and the cosmic X-ray background, γ-ray bursts (GRBs), and other energetic phenomena. Then, over the past 15 years or so, we saw a great progress powered by the advent of solid-state detectors (CCDs, IR arrays, bolometers, etc.), and cheap and ubiquitous computing, with discoveries of extrasolar planets, brown dwarfs, young and forming galaxies at high redshifts, the cosmic acceleration (the "dark energy"), the solution of the mystery of GRBs, and so on. We are now witnessing the next phase of the IT revolution, which will likely lead to another golden age of discovery in astronomy. VO is the framework to effect this process.

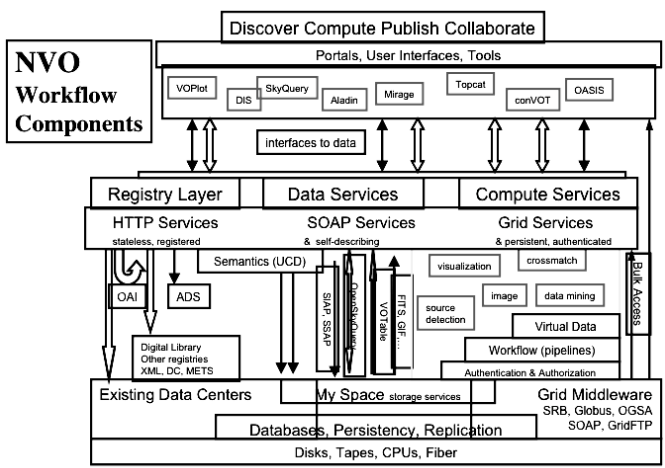

**Figure 3.** A schematic VO structure from a system architect's viewpoint. Much of the development so far has centered on the development of data and communications infrastructure. However, the principal scientific returns will come as the top layers get more developed, with a variety of DM/KDD tools applied on the great data sets already in hand, as well as the forthcoming ones. Figure courtesy of Roy Williams.

In physical sciences, and astronomy in particular, new discoveries can be classed as conceptual (e.g., the quantum theory, relativity, strings/branes, etc.), or as phenomenological (e.g., quasars, cosmic backgrounds, dark matter and dark energy, etc.). Phenomenological discoveries are usually made either by opening a new domain of the parameter space (e.g., radio astronomy, X-ray astronomy, etc.), by pushing further along some axis of the observable parameter space (e.g., deeper in flux, higher in angular or temporal resolution, etc.), by expanding the coverage of the parameter space and thus finding rare types of objects or phenomena which would be missed in sparse observations, or by making connections between different types of observations (for example, optical identification of radio sources leading to the discovery of quasars). In a more steady mode of research, application of well understood physics, constrained by observations, leads to understanding of various astronomical objects and phenomena; e.g., stellar structure and evolution.

This implies two kinds of discovery strategies: covering a large volume of the parameter space, with many sources, measurements, etc., as is done very well by massive sky surveys; and connecting as many different types of observations as possible (e.g., in a multi-wavelength, multi-epoch, or multi-scale manner), so that the potential for discovery increases as the number of connections, i.e., as the number of the federated data sets, squared. Both approaches are naturally suited for the VO.

### III. Some General Problems and Challenges

There are many non-trivial technological and methodological problems posed by the challenges of data abundance. We note two important trends, which seem to particularly distinguish the new, information-rich science from the past:

(1) *Most data will never be seen by humans*. This is a novel experience for scientists, but the sheer volume of TB-scale data sets (or larger) makes it impractical to do even a most cursory examination of all data. This implies a need for reliable data storage, networking, and database-related technologies, standars, and protocols. The problem of automated data quality control is particularly significant, and may require AI-based tools.

(2) *Most data and data constructs, and patterns present in them, cannot be comprehended by humans directly*. This is a direct consequence of a growth in complexity of information, mainly its multidimensionality. This requires the use or development of novel data mining (DM) or knowledge discovery in databases (KDD) and data understanding (DU) technologies, hyperdimensional visualization, etc. The use of AI/machine-assisted discovery may become a standard scientific practice.

This is where the qualitative differences in the way science is done in the 21$^{st}$ century will come from; the changes are not just quantitative, based on the data volumes alone.

Thus, a modern scientific discovery process can be outlined as follows:

*Data gathering*: raw data streams produced by various measuring devices. Instrumental effects are removed and calibrations applied in the domain-specific manner, usually through some data reduction pipeline (DRP). Depending on the complexity of data and sources of noise, use of automated, machine-learning tools can be very useful at this stage.

*Data farming*: storage and archiving of the raw and processed data, metadata, and derived data products, including issues of optimal database architectures, indexing, searchability, interoperability, data fusion, etc. While much remains to be done, these challenges seem to be fairly well understood, and much progress is being made.

*Data mining*: including clustering analysis, automated classification, outlier or anomaly searches, pattern recognition, multivariate correlation searches, and scientific visualization, all of them usually in some high-dimensional parameter space of measured attributes or imagery. This is where the key technical challenges are now. These are the tools which must be developed for the new, information-rich science of the 21$^{st}$ century. It is likely that many of them would have to include some form of machine intelligence.

And finally, we reach the as-yet murky area of *data*

*understanding*, leading to the actual new knowledge. The problems here are essentially methodological in nature. We need to learn how to ask new types of questions, enabled by the increases in the data volume, complexity, and quality, and the advances provided by IT. This is where the scientific creativity comes in. However, we may be making machine-assisted discoveries, driven by the extreme complexity of data constructs, and the needs to visualize hyperdimensional spaces; human intuition is closely coupled to our ability to visualize the processes and phenomena we study.

subset of dimensions (e.g., one could imagine a significant 65-variate correlation embedded in a 312-dimensional parameter space – and then face the challenge of interpreting it!). A crude illustration of the clustering problem is shown in Fig. 4, and a real-life example (from a mere 2-dimensional parameter space) in Fig. 5.

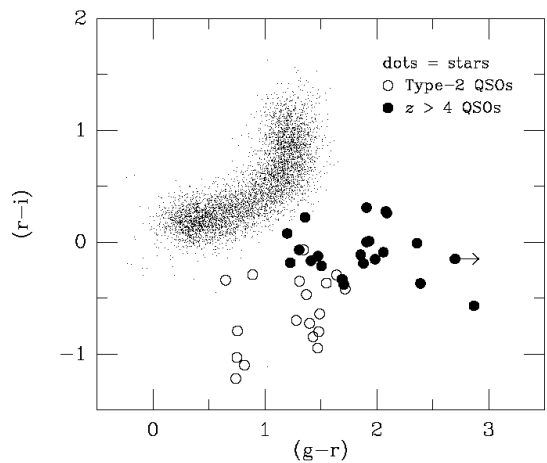

**Figure 5.** A simple example of clustering in a parameter space as a means to astrophysical discovery, from the real data collected in the DPOSS survey (see [12,13,14] for more details). What is show are two colors (green to red, and red to infrared) derived from the survey measurements. Only objects morphologically classified as stars (i.e., PSF like in the images) are used. The dots represent normal Galactic stars. The solid circles are high-redshift quasars, and the open circles are the so-called Type 2 quasars, and both classes are rare and astrophysically interesting. They are clearly separated from the locus of stars in this parameter space. Analysis of highly-multidimensional parameter spaces from individual or federated sky surveys promises to yield more such discoveries, and should become a standard technique in a VO.

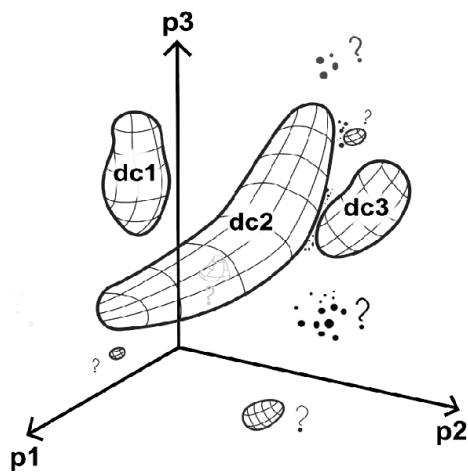

**Figure 4.** A schematic illustration of the problem of clustering analysis in some parameter space. In this example, there are 3 dimensions, p1, p2, and p3 (e.g., some flux ratios or morphological parameters), and most of the data points belong to 3 major clusters, denoted dc1, dc2, and dc3 (e.g., stars, galaxies, and ordinary quasars). One approach is to isolate these major classes of objects for some statistical studies, e..g, stars as probes of the Galactic structure, or galaxies as probes of the large scale structure of the universe, and filter out the "anomalous" objects. A complementary view is to look for other, less populated, but statistically significant, distinct clusters of data points, or even individual outliers, as possible examples of rare or unknown types of objects. Another possibility is to look for holes (negative clusters) within the major clusters, as they may point to some interesting physical phenomenon, or to a problem with the data.

To give some specific examples of challenges ahead, let us consider the general area of exploration of observable parameter spaces, which would be a typical VO activity in exploring the massive sky surveys and their federation, and clustering analysis in particular [12–16]. Generally, original image data are processed and catalogs of detected sources are derived, and many parameters (attributes) measured for each source. A typical survey may detect ~ $10^9$ sources and provide measurements or upper limits for ~ $10^2 – 10^3$ parameters for each one. The problem thus becomes one of characterizing the distribution of N ~ $10^9$ data vectors in a parameter space of D ~ $10^2 – 10^3$ dimensions. This forms a highly non-trivial problem setup for clustering analysis.

Typical questions one may ask include: How many distinct types of objects are there, and what are the classification probabilities for each object and class? What are the outliers (possibly rare new types of objects)? Are there significant multivariate correlations present in the data, possibly in some

The first technical challenge comes from the size and dimensionality of the data sets, i.e., the scalability of the clustering and classification (or indeed any other DM/KDD) algorithms. For the number of data vectors N, in an D-dimensional parameter space, most clustering or correlations algorithms scale as N log N, or even $N^2$, and as $D^2$, or even steeper. Bayesian and likelihood methods tend to scale even more steeply, as $N^m$, where m > 3, and $D^k$, where k > 1. Clustering analysis then becomes a computationally prohibitively expensive problem, especially driven by the hyperdimensionality of these data sets. Computational brute force is not always a practical or even a viable solution; better algorithms are needed. The problem is even sharper if one requires active, real-time (or time-critical) data mining in massive data streams, as expected, e.g., in the synoptic sky surveys.

One technique which can simplify the problem of data volume and hyperdimensionality is the multi-resolution clustering. In this regime, parameters expensive to estimate, such as the number of classes and the initial broad clustering are quickly estimated using traditional techniques, and then one could proceed to refine the model locally and globally by iterating until some objective statistical (e.g., Bayesian) criterion is satisfied. One can also use intelligent sampling methods where one forms prototypes of the case vectors and thus reduces the number of cases to process. Prototypes can be determined from simple algorithms to get a rough estimate, and then refined using more sophisticated techniques. A clustering algorithm can operate in prototype space. The clusters found can later refined by locally replacing each prototype by its constituent population and reanalyzing the cluster. Various techniques for dimensionality reduction, including principal component analysis and others can be used as preprocessing techniques to automatically derive the dimensions that contain most of the relevant information.

Aside from the computational challenges with large numbers of data vectors and a large dimensionality, this poses some highly non-trivial statistical problems. The problems are driven not just by the size of the data sets, but mainly by the heterogeneity and intrinsic complexity of the data.

For example, some of the source parameters would be primary measurements, and others may be derived attributes, such as the star-galaxy classification, some may be flags rather than numbers, some would have error-bars associated with them, and some would not, and the error-bars may be functions of some of the parameters, e.g., fluxes. Some measurements would be present only as upper or lower limits. Some would be affected by glitches due to instrumental problems, and if a data set consists of a merger of two or more surveys, e.g., cross-matched optical, infrared, and radio (and this would be a common scenario within a VO), then some sources would be misidentified, and thus represent erroneous combinations of subsets of data dimensions. Surveys would be also affected by selection effects operating explicitly on some parameters (e.g., coordinate ranges, flux limits, etc.), but also mapping onto some other data dimensions through correlations of these properties; some selection effects may be unknown.

Additional complications may derive from the intrinsic nature of clustering and distributions present in the data. For example: The object classes form multivariate "clouds" in the parameter space, but these clouds in general need not be Gaussian: some may have a power-law or exponential tails in some or all of the dimensions, and some may have sharp cutoffs, etc. The clouds may be well separated in some of the dimensions, but not in others. How can we objectively decide which dimensions are irrelevant, and which ones are useful? The topology of clustering may not be simple: there may be clusters within clusters, holes in the data distribution (negative clusters?), multiply-connected clusters, etc.

Perhaps the first methodological choice in this type of exploration is the question of supervised vs. unsupervised classification. If the number of object classes $k$ is known (or declared) a priori, and training data set of representative objects is available, the problem reduces to supervised classification, where tools such as Artificial Neural Nets or Decision Trees can be used. Searches for known types of objects with predictable signatures in the parameter space (e.g., high-redshift quasars, in the example shown in Fig. 5) can be also cast in this way.

However, a more interesting and less biased approach is where the number of classes $k$ is not known, and it has to be derived from the data themselves. This opens a possibility of discovery of genuinely new types of astronomical objects or phenomena. The problem of unsupervised classification is to determine this number in some objective and statistically sound manner, and then to associate class membership probabilities for all objects. Majority of objects may fall into a small number of classes, e.g., normal stars or galaxies. What is of a special interest are objects which belong to much less populated clusters, or even individual outliers with low membership probabilities for any major class.

There is a history of use of supervised and unsupervised classifiers in astronomy, primarily for automated star-galaxy separation in digital sky surveys, e.g., [17-23], but a full-scale application to major digital sky surveys, aside frim the star-galaxy classification in visible-light surveys, yet remains to be done. We can expect a much broader use of such techniques in the future.

Given this computational and statistical complexity, blind applications of the commonly used (commercial or home-brewed) clustering algorithms could produce some seriously misleading or simply wrong results. The clustering methodology must be robust enough to cope with these problems, and the outcome of the analysis must have a solid statistical foundation. It is then fair to say that adequate toolbox of properly scalable DM/KDD algorithms for analysis of such massive and complex data sets simply does not yet exist. This is an example of a potentially great synergy between domain scientists (in this case astronomers) and information and computer scientists and statisticians.

Another key issue is interoperability and reusability of algorithms and models in a wide variety of problems posed by a rich data environment such as federated digital sky surveys in a VO. Implementation of clustering analysis algorithms must be done with this in mind.

In many situations, scientifically informed input is needed in designing the clustering experiments. Some observed parameters may have a highly significant, large dynamical range, dominate the sample variance, and naturally invite division into clusters along the corresponding parameter axes; yet they may be completely irrelevant or uninteresting scientifically. Design and application of clustering algorithms should be based on a close, working collaboration between astronomers and computer scientists and statisticians. There are too many unspoken assumptions, historical background knowledge specific to the given discipline, and opaque jargon; constant communication and interchange of ideas are essential.

The second, and perhaps even more critical part of the "curse of hyperdimensionality" is the visualization of these highly-dimensional data parameter spaces. Humans are biologically limited to visualize patterns and scenes in 2 or 3 dimensions, and while some clever tricks have been developed to increase the maximum visualizable dimensions, in practice it is hard to push much beyond D = 4 or 5. Mathematically, we understand the meaning of clustering and correlations in an arbitrary number of parameter space dimensions, but how can we actually visualize such structures? Yet, recognizing and intuitively comprehending such complex data constructs may lead to some crucial new astrophysical insights. This is an essential part of the intuitive process of scientific discovery, and critical to data understanding and knowledge extraction.

Effective and powerful data visualization, applied in the parameter space itself, must be an essential part of the interactive clustering analysis. Good visualization tools are also critical for the interpretation of results, especially in an iterative environment. While clustering algorithms can assist in the partitioning of the data space, and can draw the attention to anomalous objects, ultimately a scientist guides the experiment and draws the conclusions. This may be another area where AI-assisted exploration of complex data sets would become a necessary part of a scientist's toolkit.

## IV. THE EVOLVING ROLE OF SCIENTIFIC COMPUTING

The modern scientific methodology originated in the 17$^{th}$ century, and a healthy interplay of analytical and experimental work has been driving the scientific progress ever since. But then, in mid-20$^{th}$ century, something new came along: computing as a new way of doing science, primarily through numerical simulations of phenomena too complex to be analytically tractable. Simulations are thus more than just a substitute for analytical theory: there are many phenomena in the physical universe where simulations (incorporating, of course, the right physics and equations of motion) are the *only* way in which some phenomena can be described and predicted. Recall that even the simplest Newtonian mechanics can solve exactly only a 2-body problem; for $N \geq 3$, numerical solutions are necessary. Other examples in astronomy include star and galaxy formation, dynamics and evolution of galaxies and large-scale structure, stellar explosions, anything involving turbulence, etc. Simulations relate, can stimulate, or be explained by both analytical theory and experiments or observations.

While numerical simulations and other computational means of solving complex systems of equations continue to thrive, there is now a new and growing role of scientific computing, which is data-driven.

Data- or information-driven computing, which spans all of the aspects of a modern scientific work described above, and more, is now becoming the dominant form of scientific computing, and an essential component of gathering, storing, preserving, accessing, and – most of all – analyzing massive amounts of complex data, and extracting knowledge from them. It is fundamentally changing the way in which science is done in the 21$^{st}$ century.

It is an interesting epistemological question whether this advent of information and computing intensive science represents something qualitatively new, on par with the traditional analytical theory and experiments (i.e., quantitative measurements), or numerical theory, or is it just an extension of the traditional experimental/observational work, which simply uses IT as a tool. At what point is the quantitative change, brought by a many orders of magnitude increase in the data volumes and complexity, becoming a qualitative one? This is hard to answer, since the new methodology for the information-rich science is still in its nascent state.

Another, mildly provocative idea, is that applied computer science is now playing the role which mathematics did from the 17$^{th}$ through the 20$^{th}$ centuries: providing an orderly, formal framework and exploratory apparatus for other sciences. Aside from its apparently happy affair with the string theory, it is hard to tell what mathematics is doing for other sciences today; most of the mathematics scientists use today was developed over a century ago.

There are also sociological changes afoot. As the data become ever cheaper, more available, and easily accessible, the focus of values will shift from the ownership of data (or instruments used to gather them) to the ownership of expertise and ideas. Computationally astute scientists will grow from a status of nearly second-class citizens in the academic pantheon today, to become the dominant experts in every field.

Computationally driven and enabled science also plays another, very important societal role: it is empowering an unprecedented pool of talent. With distributed scientific frameworks like VO, which provide open access to data and tools for their exploration, anyone, anywhere, with a decent internet connection can do a first rate science, learn about what others area doing, and communicate their results. This should be a major boon for countries without expensive scientific facilities, and individuals at small or isolated institutions. The human talent is distributed geographically much more broadly than money or other resources. By broadening access to anyone with good ideas and good work habits, science will prosper at a much faster rate that it has done historically, with all of the subsequent societal benefits that implies.

The web is of course a supreme public outreach medium. As the science migrates to web-based research environments such as the VO, public outreach and education at all levels, from pre-school to graduate school, will discover some powerful "weapons of mass instruction".

## V. CONCLUDING REMARKS

As the preceding discussion implies, the critical challenges in scientific computing are now in the arena of software (broadly speaking): from various database and data farming tasks, to DM/KDD, and data exploration and understanding. Computing hardware is to a first approximation becoming infinitely powerful and infinitely cheap (yet somehow hardware producers continue to make money!). On the other hand, the beneficial exponential improvement in price/performance does not apply in the world of software and algorithm development. Therefore, the most valuable long-term investments will be on the side of software and methodology.

The key technological challenges today are in the development of efficiently scalable DM/KDD algorithms, hyperdymensional visualization, and other techniques for data exploration. Equally important are challenges of effective visualization of highly hyperdimensional spaces and data constructs. Such techniques are necessary in order to make the full use of the rich information content of massive and complex data sets.

However, as the visualization problem clearly shows, we may be reaching limits of human intuitive comprehension of extremely complex data structures and constructs, either from the measurements or from numerical simulations. Use of machine learning and AI techniques may become an essential part of a scientific methodology. Just as computers now help us handle numbers, text, images, etc., on scales and at speeds well beyond unaided human ability, software machinery of the future may help us grasp and understand complexities which are beyond our reach now.

Just as technology derives from a progress in science, progress in science, especially experimental/observational, is driven by the progress in technology. This positive feedback loop will continue, as the IT revolution unfolds. Practical CS/IT solutions cannot be developed in a vacuum; having real-life testbeds, and functionality driven by specific application demands is essential. Recall that the WWW originated as a scientific application, and noone could have predicted its ultimate impact at the time. Today, grid technology is being developed by physicists, astronomers, and other scientists. The needs of information-driven science are broadly applicable to information-intensive economy in general, as well as other domains (entertainment, media, security, education, etc.). Who knows what world-changing technology, perhaps even on par with the WWW itself, would emerge from the synergy of computationally enabled science, and science-driven information technology?

ACKNOWLEDGMENT

The author thanks numerous colleagues who helped develop the Virtual Observatory concept, and many of the ideas described in this paper. In particular, special thanks are due to Alex Szalay, Roy Williams, Ashish Mahabal, Matthew Graham, Robert Brunner, Jim Gray, Tom Prince, Roy Gal, Reinaldo de Carvalho, Nick Weir, Usama Fayyad, Joe Jacob, Bob Hanisch, Dave De Young, and many others. Joe Bredekamp of NASA provided an essential early support and encouragement. Some of the ideas described here were developed while the author was enjoying the stimulating intellectual environment of the Aspen Center for Physics.

**S. G. Djorgovski** is a Professor of Astronomy and a Director of the Center for Advanced Computing Research at Caltech. He is an author or coauthor of several hundred scientific publications, covering a broad range of astronomical discoveries and other results, and one of the founders of the Virtual Observatory concept. He got his PhD in Berkeley in 1985, and was a Harvard Junior Fellow before joining the Caltech faculty in 1987. He was a Presidential Young Investigator, and an Alfred P. Sloan Foundation Fellow, among other distinctions.